\documentclass{IEEEtran}
\IEEEoverridecommandlockouts

\makeatletter

\newcommand{\multiline}[1]{%
  \begin{tabularx}{\dimexpr\linewidth-\ALG@thistlm}[t]{@{}X@{}}
    #1
  \end{tabularx}
}

\def\endthebibliography{%
  \def\@noitemerr{\@latex@warning{Empty `thebibliography' environment}}%
  \endlist
}
\makeatother

\usepackage[cmex10]{amsmath}
\usepackage{amssymb,amsthm}
\DeclareMathOperator*{\argmax}{arg\,max}

\usepackage{bm}
\usepackage{amsfonts}
\usepackage{relsize}
\usepackage{cuted}
\usepackage{subfig}
\usepackage{graphicx}
\usepackage{epstopdf}
\usepackage{algorithm,tabularx}
\usepackage[noend]{algpseudocode}
\usepackage{bbm}
\usepackage{comment}
\usepackage{multirow}
\usepackage{pbox}
\usepackage{array}
\usepackage{booktabs}
\usepackage{multicol}
\usepackage{url}
\usepackage{hyperref}
\usepackage{cleveref}

\newtheorem{theorem}{Theorem}

\newcommand*\diff{\mathop{}\!\mathrm{d}}

\usepackage{textcomp}
\usepackage{cite}
\usepackage{subfig}
\usepackage{xcolor}
\def\BibTeX{{\rm B\kern-.05em{\sc i\kern-.025em b}\kern-.08em
    T\kern-.1667em\lower.7ex\hbox{E}\kern-.125emX}}
\usepackage{enumitem} % per enumerate con resume
\usepackage[font=small,labelfont=bf]{caption}
\usepackage{capt-of}
\algnewcommand{\Inputs}[1]{%
  \State \textbf{Inputs:}
  \Statex \hspace*{\algorithmicindent}\parbox[t]{.8\linewidth}{\raggedright #1}
}
\algnewcommand{\Initialize}[1]{%
  \State \textbf{Initialize:}
  \Statex \hspace*{\algorithmicindent}\parbox[t]{.8\linewidth}{\raggedright #1}
}

\makeatletter
\def\BState{\State\hskip-\ALG@thistlm}
\makeatother
\title{Cooperative Channel Capacity Learning}

\author{\IEEEauthorblockN{Nunzio A. Letizia, Andrea M. Tonello, \IEEEmembership{Senior Member, IEEE}, and H. Vincent Poor, \IEEEmembership{Life Fellow, IEEE}} 
\thanks{The work of N. A. Letizia was supported in part by the mobility funding scheme of the University of Klagenfurt. \\
The work of H. V. Poor was supported in part by the U.S National Science Foundation under Grants CCF-1908308 and CNS-2128448. \\
N. A. Letizia and A. M. Tonello are with Universit\"{a}t Klagenfurt, Institute of Networked and Embedded Systems, Klagenfurt, 9020, Austria, e-mail: nunzio.letizia@aau.at, andrea.tonello@aau.at \\
H. V. Poor is with the Department of Electrical and Computer Engineering, Princeton University, Princeton, NJ 08544, USA, e-mail:  poor@princeton.edu}
}


\IEEEoverridecommandlockouts
\begin{document}

\maketitle
%\thispagestyle{plain}

%\pagestyle{plain} for number pages

\begin{abstract}
In this paper, the problem of determining the capacity of a communication channel is formulated as a cooperative game, between a generator and a discriminator, that is solved via deep learning techniques. The task of the generator is to produce channel input samples for which the discriminator ideally distinguishes conditional from unconditional channel output samples.
%paired and unpaired channel input-output samples. 
The learning approach, referred to as cooperative channel capacity learning (CORTICAL), provides both the optimal input signal distribution and the channel capacity estimate. Numerical results demonstrate that the proposed framework learns the capacity-achieving input distribution under challenging non-Shannon settings.
\end{abstract}

\begin{IEEEkeywords}
channel capacity, capacity-achieving distribution, deep learning, capacity learning. 
\end{IEEEkeywords}

\section{Introduction}
\label{sec:introduction}
\begin{comment}
Nonetheless, lack of performance guarantees and strong dependence on architecture and training parameters inhibit the adoption of deep learning.  
\end{comment}

Data-driven models for physical layer communications \cite{Oshea2017} have recently seen a surge of interest. While the benefit of a deep learning (DL) approach is evident for tasks like signal detection \cite{Ye2017}, decoding \cite{MIND2022} and channel estimation \cite{Ye2018,Soltani2019}, the fundamental problem of estimating the capacity of a channel remains elusive. Despite recent attempts via neural mutual information estimation \cite{Aharoni2020, Letizia2021,Farhad2022}, it is not clear yet whether DL can provide novel insights.

For a discrete-time continuous memoryless vector channel, the capacity is defined as
\begin{equation}
C = \max_{p_X(\mathbf{x})} I(X;Y),
\end{equation}
where $p_X(\mathbf{x})$ is the input signal probability density function (PDF), $X$ and $Y$ are the channel input and output random vectors, respectively, and $I(X; Y)$ is the mutual information (MI) between $X$ and $Y$. 
The channel capacity problem involves both determining the capacity-achieving distribution and evaluating the maximum achievable rate. Only a few special cases, e.g., additive noise channels with specific noise distributions under input power constraints, have been solved so far. When the channel is not an additive noise channel, analytical approaches become mostly intractable leading to numerical solutions, relaxations \cite{CuttingPlane}, capacity lower and upper bounds \cite{McKellips2004}, and considerations on the support of the capacity-achieving distribution \cite{Smith1971, Dytso2020}. It is known that the capacity of a discrete memoryless channel can be computed using the Blahut-Arimoto (BA) algorithm \cite{BlahutArimoto}, whereas a particle-based BA method was proposed in \cite{Dauwels2005} to tackle the continuous case although it fails to scale to high-dimension vector channels.

In the following section, we show that a data-driven approach can be pursued to obtain the channel capacity. In particular, we propose to learn the  capacity and the capacity-achieving distribution of any discrete-time continuous memoryless vector channel via a cooperative framework referred to as CORTICAL. The framework is inspired by generative adversarial networks (GANs) \cite{Goodfellow2014} but it can be interpreted as a dual version using an appropriately defined value function. In fact, CORTICAL comprises two blocks cooperating with each other: a generator that learns to sample from the capacity-achieving distribution, and a discriminator that learns to differentiate paired channel input-output samples from unpaired ones, i.e. it distinguishes the joint PDF $p_{XY}(\mathbf{x},\mathbf{y})$ from the product of marginal $p_{X}(\mathbf{x})p_{Y}(\mathbf{y})$, so as to estimate the MI. 

The paper is organized as follows: Sec. \ref{sec:cortical} describes the main idea and contribution. The parametric implementation and training algorithm are discussed in Sec. \ref{sec:implementation}. Sec. \ref{sec:results} presents numerical results for different types of channels. Sec. \ref{sec:conclusions} concludes the paper.

\section{Cooperative Principle for Capacity Learning}

\label{sec:cortical}
\begin{figure*}[h]
	\centering
	\includegraphics[scale=0.40]{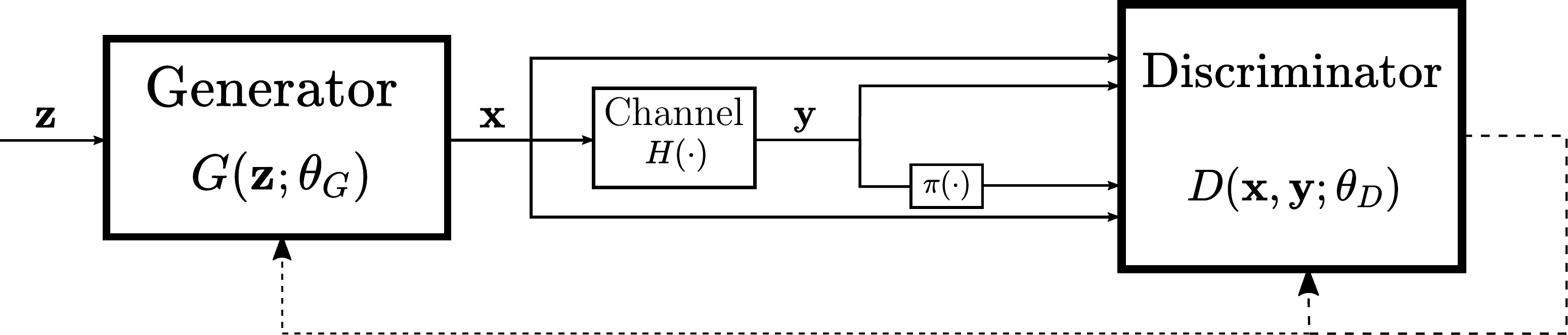}
	\caption{CORTICAL, Cooperative framework for capacity learning: a generator produces input samples with distribution $p_X(\mathbf{x})$ and a discriminator attempts to distinguish between paired and unpaired channel input-output samples.}
	\label{fig:Cooperative_networks}
\end{figure*} 
To understand the working principle of CORTICAL and why it can be interpreted as a cooperative approach, it is useful to briefly review how GANs operate. 

In the GAN framework, the adversarial training procedure for the generator $G$ consists of maximizing the probability of the discriminator $D$ making a mistake. If $\mathbf{x}\sim p_{\text{data}}(\mathbf{x})$ are the data samples and $\mathbf{z}\sim p_{Z}(\mathbf{z})$ are the latent samples, the Nash equilibrium is reached when the minimization over $G$ and the maximization over $D$ of the value function
\begin{align}
\mathcal{V}(G,D) = \; & \mathbb{E}_{\mathbf{x} \sim p_{\text{data}}(\mathbf{x})}\biggl[\log \bigl(D(\mathbf{x})\bigr)\biggr] \nonumber \\
& +\mathbb{E}_{\mathbf{z} \sim p_{Z}(\mathbf{z})}\biggl[\log\bigl(1-D\bigl(G(\mathbf{z})\bigr)\bigr)\biggr],
\label{eq:GAN_value_function}
\end{align}
is attained so that $G(\mathbf{z})\sim p_{\text{data}}(\mathbf{x})$. Concisely, the minimization over $G$ forces the generator to implicitly learn the distribution that minimizes the given statistical distance. 
%$D=\frac{1}{2}$ Cioe' l'outpit di D e`1/2 ? 

Conversely to GANs, the channel capacity estimation problem requires the generator to learn the distribution maximizing the mutual information measure. 
Therefore, the generator and discriminator need to play a cooperative max-max game with the objective for $G$ to produce channel input samples for which $D$ exhibits the best performance in distinguishing (in the Kullback-Leibler sense) paired and unpaired channel input-output samples. Thus, the discriminator of CORTICAL is fed with both the generator and channel's output samples, $\mathbf{x}$ and $\mathbf{y}$, respectively.
Fig.\ref{fig:Cooperative_networks} illustrates the proposed cooperative framework that learns both for the cooperative training, as discussed next.

\begin{theorem}
\label{theorem:Theorem1}
 Let $X\sim p_X(\mathbf{x})$ and $Y|X\sim p_{Y|X}(\mathbf{y}|\mathbf{x})$ be the vector channel input and conditional output, respectively. Let $Y = H(X)$ with $H(\cdot)$ being the stochastic channel model and let $\pi(\cdot)$ be a permutation function \footnote{The permutation takes place over the temporal realizations of the vector $Y$ so that $\pi(Y)$ and $X$ can be considered statistically independent vectors.} over the realizations of $Y$ such that $p_{\pi(Y)}(\pi(\mathbf{y})|\mathbf{x}) = p_{Y}(\mathbf{y})$. In addition, let $G$ and $D$ be two functions in the non-parametric limit such that $X = G(Z)$ with $Z\sim p_Z(\mathbf{z})$ being a latent random vector. If $\mathcal{J}_{\alpha}(G,D)$, $\alpha>0$, is the value function defined as 
\begin{align}
\mathcal{J}_{\alpha}(G,D) = \; & \alpha \cdot \mathbb{E}_{\mathbf{z} \sim p_{Z}(\mathbf{z})}\biggl[\log \biggl(D\biggl(G(\mathbf{z}),H(G(\mathbf{z}))\biggr)\biggr)\biggr] \nonumber \\
& -\mathbb{E}_{\mathbf{z} \sim p_{Z}(\mathbf{z})}\biggl[D\biggl(G(\mathbf{z}),\pi(H(G(\mathbf{z})))\biggr)\biggr],
\label{eq:value_function}
\end{align}
then the channel capacity $C$ is the solution of
\begin{equation}
C = \max_{G} \max_{D} \frac{\mathcal{J}_{\alpha}(G,D)}{\alpha} + 1- \log(\alpha),
\end{equation}
and $\mathbf{x}=G^*(\mathbf{z})$ are samples from the capacity-achieving distribution, where 
\begin{equation}
G^* = \argmax_G \max_D \mathcal{J}_{\alpha}(G,D).
\end{equation}
\end{theorem}

\begin{proof}
The value function in (\ref{eq:GAN_value_function}) can be written as
\begin{align}
\mathcal{J}_{\alpha}(G,D) = \; & \alpha \cdot \mathbb{E}_{(\mathbf{x},\mathbf{y}) \sim p_{XY}(\mathbf{x},\mathbf{y})}\biggl[\log \biggl(D(\mathbf{x},\mathbf{y})\biggr)\biggr] \nonumber \\
& -\mathbb{E}_{(\mathbf{x},\mathbf{y}) \sim p_{X}(\mathbf{x}) p_Y(\mathbf{y})}\biggl[D(\mathbf{x},\mathbf{y})\biggr].
\label{eq:value_function_p}
\end{align}
Given a generator $G$ that maps the latent (noise) vector $Z$ into $X$, we need firstly to prove that $\mathcal{J}_{\alpha}(G,D)$ is maximized for 
\begin{equation}
D(\mathbf{x},\mathbf{y})=D^*(\mathbf{x},\mathbf{y})=\alpha  \frac{p_{XY}(\mathbf{x},\mathbf{y})}{p_{X}(\mathbf{x}) p_Y(\mathbf{y})}.
\end{equation}
Using the Lebesgue integral to compute the expectation
\begin{align}
\mathcal{J}_{\alpha}(G, D) = \; & \alpha  \int_{\mathbf{y}} \int_{\mathbf{x}}\biggl[p_{XY}(\mathbf{x},\mathbf{y}) \log \biggl(D(\mathbf{x},\mathbf{y})\biggr) \nonumber \\ 
& - p_{X}(\mathbf{x}) p_Y(\mathbf{y}) \biggl(D(\mathbf{x},\mathbf{y})\biggr)\biggr] \diff \mathbf{x} \diff \mathbf{y},
\label{eq:value_function_q}
\end{align}
taking the derivative of the integrand with respect to $D$ and setting it to $0$, yields the following equation in $D$:
\begin{equation}
\alpha  \frac{p_{XY}(\mathbf{x},\mathbf{y})}{D(\mathbf{x},\mathbf{y})} - p_{X}(\mathbf{x}) p_Y(\mathbf{y})=0,
\end{equation}
whose solution is the optimum discriminator
\begin{equation}
D^*(\mathbf{x},\mathbf{y}) = \alpha  \frac{p_{XY}(\mathbf{x},\mathbf{y})}{p_{X}(\mathbf{x}) p_Y(\mathbf{y})}.
\end{equation}
In particular, $\mathcal{J}_{\alpha}(D^*)$ is a maximum since the second derivative of the integrand $-\alpha  \frac{p_{XY}(\mathbf{x},\mathbf{y})}{D^2(\mathbf{x},\mathbf{y})}$ is a non-positive function.
Therefore, substituting $D^*(\mathbf{x},\mathbf{y})$ in \eqref{eq:value_function_q} yields
%\begin{align}
%\mathcal{J}_{\alpha}(G,D^*) =  \; \alpha & \cdot \mathbb{E}_{(\mathbf{x},\mathbf{y}) \sim p_{XY}(\mathbf{x},\mathbf{y})}\biggl[\log \biggl(\alpha  \frac{p_{XY}(\mathbf{x},\mathbf{y})}{p_{X}(\mathbf{x}) p_Y(\mathbf{y})} \biggr)\biggr] \nonumber \\
%& -\mathbb{E}_{(\mathbf{x},\mathbf{y}) \sim p_{X}(\mathbf{x}) p_Y(\mathbf{y})}\biggl[\alpha  \frac{p_{XY}(\mathbf{x},\mathbf{y})}{p_{X}(\mathbf{x}) p_Y(\mathbf{y})}\biggr],
%\end{align}
%and using the Lebesgue integral to compute the expectation
\begin{align}
\mathcal{J}_{\alpha}(G,D^*) = \; & \alpha \int_{\mathbf{y}} \int_{\mathbf{x}}{\biggl[p_{XY}(\mathbf{x},\mathbf{y}) \log \biggl(\alpha  \frac{p_{XY}(\mathbf{x},\mathbf{y})}{p_{X}(\mathbf{x}) p_Y(\mathbf{y})}} \biggr) \nonumber \\
& - p_{X}(\mathbf{x}) p_Y(\mathbf{y}) \biggl(\alpha \frac{p_{XY}(\mathbf{x},\mathbf{y})}{p_{X}(\mathbf{x}) p_Y(\mathbf{y})}\biggr)\biggr] \diff \mathbf{x} \diff \mathbf{y}.
\end{align}
Now, it is simple to recognize that the second term on the right hand side of the above equation is equal to $-\alpha$. Thus, 
\begin{align}
\mathcal{J}_{\alpha}(G,D^*) & = \; \alpha \cdot \mathbb{E}_{(\mathbf{x},\mathbf{y}) \sim p_{XY}(\mathbf{x},\mathbf{y})}\biggl[\log \biggl(\frac{p_{XY}(\mathbf{x},\mathbf{y})}{p_{X}(\mathbf{x}) p_Y(\mathbf{y})} \biggr)\biggr] \nonumber \\
& + \alpha \log(\alpha) - \alpha \nonumber \\
& = \alpha \bigl(I(X;Y)+\log(\alpha)-1\bigr).
\end{align}
Finally, the maximization over the generator $G$ results in the mutual information maximization since $\alpha$ is a positive constant. We therefore obtain,  
%\begin{equation}
%\max_{G} \mathcal{J}_{\alpha}(G,D^*) = \max_{G} %\alpha \bigl( \log(\alpha) - 1 + {I}(X;Y)\bigr).
%\end{equation}
%Hence, since $G$ models the distribution $p_X(\mathbf{x})$ and $\alpha$ is a positive constant,
\begin{equation}
\max_{G} \mathcal{J}_{\alpha}(G,D^*)  + \alpha - \alpha \log(\alpha) = \alpha \max_{p_X(\mathbf{x})} {I}(X;Y).
\end{equation}
\end{proof}

Theorem \ref{theorem:Theorem1} states that at the equilibrium, the generator of CORTICAL has implicitly learned the capacity-achieving distribution with the mapping $\mathbf{x}=G^*(\mathbf{z})$. 
In contrast to the BA algorithm, here the generator samples directly from the optimal input distribution $p_X(\mathbf{x})$ rather than explicitly modelling it. No assumptions on the input distribution's nature are made. Moreover, we have access to the channel capacity directly from the value function used for training as follows:
\begin{equation}
    \label{eq:capacity}
    C = \frac{\mathcal{J}_{\alpha}(G^*,D^*)}{\alpha} + 1- \log(\alpha).
\end{equation}

In the following, we propose to parametrize $G$ and $D$ with neural networks and we explain how to train CORTICAL.
\section{Parametric Implementation}
\label{sec:implementation}
It can be shown (see Sec.4.2 in \cite{Goodfellow2014}) that the alternating training strategy described in Alg. \ref{alg:1}
converges to the optimal $G$ and $D$ under the assumption of having enough capacity and training time. Practically, instead of optimizing over the space of functions $G$ and $D$, it is reasonable to model both the generator and the discriminator with neural networks $(G,D)=(G_{\theta_G},D_{\theta_D})$ and optimize over their parameters $\theta_G$ and $\theta_D$ (see Alg. \ref{alg:1}).
We consider the distribution of the source $p_Z(\mathbf{z})$ to be a multivariate normal distribution with independent components. The function $\pi (\cdot)$ implements a random derangement of the batch $\mathbf{y}$ and it is used to obtain unpaired samples. We use simple neural network architectures and we execute $K=10$ discriminator training steps every generator training step. Details of the architecture and implementation are reported in GitHub \cite{CORTICAL_github}.
It should be noted that in Th. \ref{theorem:Theorem1}, $p_X(\mathbf{x})$ can be subject to certain constraints, e.g., peak and/or average power. Such constraints are met by the design of $G$, for instance using a batch normalization layer. 
Alternatively, we can impose peak and/or average power constraints in the estimation of capacity by adding regularization terms (hinge loss) in the generator part of the value function \eqref{eq:value_function}, as in constrained optimization problems, e.g., Sec. III of \cite{Faycal2001}. Specifically, the value function becomes
\begin{align}
& \mathcal{J}_{\alpha}(G,D) = \alpha \cdot \mathbb{E}_{\mathbf{z} \sim p_{Z}(\mathbf{z})}\biggl[\log \biggl(D\biggl(G(\mathbf{z}),H(G(\mathbf{z}))\biggr)\biggr)\biggr] \nonumber \\
& -\mathbb{E}_{\mathbf{z} \sim p_{Z}(\mathbf{z})}\biggl[D\biggl(G(\mathbf{z}),\pi(H(G(\mathbf{z})))\biggr)\biggr] \nonumber \\ 
& - \lambda_A \max(||G(\mathbf{z})||^2_2-A^2,0) - \lambda_P \max(\mathbb{E}[||G(\mathbf{z})||^2_2]-P,0),
\label{eq:value_function_lambda}
\end{align}
with $\lambda_A$ and $\lambda_P$ equal to $0$ or $1$.

\begin{algorithm}[t]
\caption{Cooperative Channel Capacity Learning}
\label{alg:1}
\begin{algorithmic}[1]
\Inputs{$N$ training steps, $K$ discriminator steps, $\alpha$.}
\For{$n=1$ to $N$}
    \For{$k=1$ to $K$}
	\State \multiline{Sample batch of $m$ noise samples $\{\mathbf{z}^{(1)},\dots,\mathbf{z}^{(m)}\}$ from $p_Z(\mathbf{z})$;}
	\State \multiline{Produce batch of $m$ channel input/output paired samples $\{(\mathbf{x}^{(1)},\mathbf{y}^{(1)}),\dots,(\mathbf{x}^{(m)},\mathbf{y}^{(m)})\}$ using the generator $G_{\theta_G}$ and the channel model $H$;}
        \State \multiline{Shuffle (derangement) $\mathbf{y}$ and get input/output unpaired samples $\{(\mathbf{x}^{(1)},\mathbf{\tilde{y}}^{(1)}),\dots,(\mathbf{x}^{(m)},\mathbf{\tilde{y}}^{(m)})\}$;}
        \State \multiline{Update the discriminator by ascending its stochastic gradient:}
        \begin{equation*}
            \nabla_{\theta_D} \frac{1}{m} \sum_{i=1}^{m}{\alpha \log\bigl(D_{\theta_D}\bigl(\mathbf{x}^{(i)},\mathbf{y}^{(i)}\bigr)\bigr)-D_{\theta_D}}\bigl(\mathbf{x}^{(i)},\mathbf{\tilde{y}}^{(i)}\bigr).
        \end{equation*}
\EndFor
    \State \multiline{Sample batch of $m$ noise samples $\{\mathbf{z}^{(1)},\dots,\mathbf{z}^{(m)}\}$ from $p_Z(\mathbf{z})$;}
    \State \multiline{Update the generator by ascending its stochastic gradient:}
        \begin{align*}
            \nabla_{\theta_G} \frac{1}{m} & \sum_{i=1}^{m}{\alpha \log\biggl( D_{\theta_D}\biggl(G_{\theta_G}\bigl(\mathbf{z}^{(i)}\bigr),H\bigl(G_{\theta_G}\bigl(\mathbf{z}^{(i)}\bigr)\bigr)\biggr)\biggr)} \\
            & -D_{\theta_D}\biggl(G_{\theta_G}\bigl(\mathbf{z}^{(i)}\bigr),\pi \bigl(H\bigl(G_{\theta_G}\bigl(\mathbf{z}^{(i)}\bigr)\bigr)\bigr)\biggr).
        \end{align*}
\EndFor
\end{algorithmic}
\end{algorithm}

\begin{comment}
For an additive white Gaussian noise (AWGN) channel under an average power constraint the exact form of the capacity and the capacity-achieving input distribution are well known. Using CORTICAL to retrieve the optimal Gaussian (of variance $P$) input is not a sufficient proof of concept since one may argue that the neural generator implements a sort of central limit theorem approximation. Therefore, we decide to apply the cooperative framework in three, more representative, non-Shannon's setups.
\end{comment}

\section{Numerical Results}
\label{sec:results}
To demonstrate the ability of CORTICAL to learn the optimal channel input distribution, we evaluate its performance in three non-standard scenarios: 1) the additive white Gaussian noise (AWGN) channel subject to a peak-power constrained input; 2) an additive non-Gaussian noise channel subject to two different input power constraints; and 3) the Rayleigh fading channel known at both the transmitter and the receiver subject to an average power constraint. These are among the few scenarios for which analysis has been carried out in the literature and thus offer a baseline to benchmark CORTICAL. For both the first and third scenarios, it is known that the capacity-achieving distribution is discrete with a finite set of mass points \cite{Smith1971,Tchamkerten2004,Dytso2020}. For the second scenario the nature of the input distribution depends on the type of input power constraint \cite{Fahs2014}. Additional results including the study of the classical AWGN channel with average power constraint are reported in GitHub \cite{CORTICAL_github}.

\subsection{Peak Power-Limited Gaussian Channels}
The capacity of the discrete-time memoryless vector Gaussian noise channel with unit-variance under peak-power constraints on the input is defined as \cite{Rassouli2016}
\begin{equation}
C(A) = \sup_{p_X(\mathbf{x}): ||X||_2\leq A} I(X;Y),
\end{equation}
where $p_X(\mathbf{x})$ is the channel input PDF and $A^2$ is the upper bound for the input signal peak power.
The AWGN Shannon capacity constitutes a trivial upper bound for $C(A)$,
\begin{equation}
C(A) \leq \frac{d}{2}\log_2\biggl(1+\frac{A^2}{d}\biggr),
\end{equation}
while a tighter upper bound for the scalar channel is provided in \cite{McKellips2004}
\begin{equation}
C(A) \leq \min\biggl\{\log_2\biggl(1+\frac{2A}{\sqrt{2\pi e}}\biggr), \frac{1}{2}\log_2\bigl(1+A^2\bigr) \biggr\}.
\end{equation}
\begin{figure}
	\centering
	\includegraphics[scale=0.22]{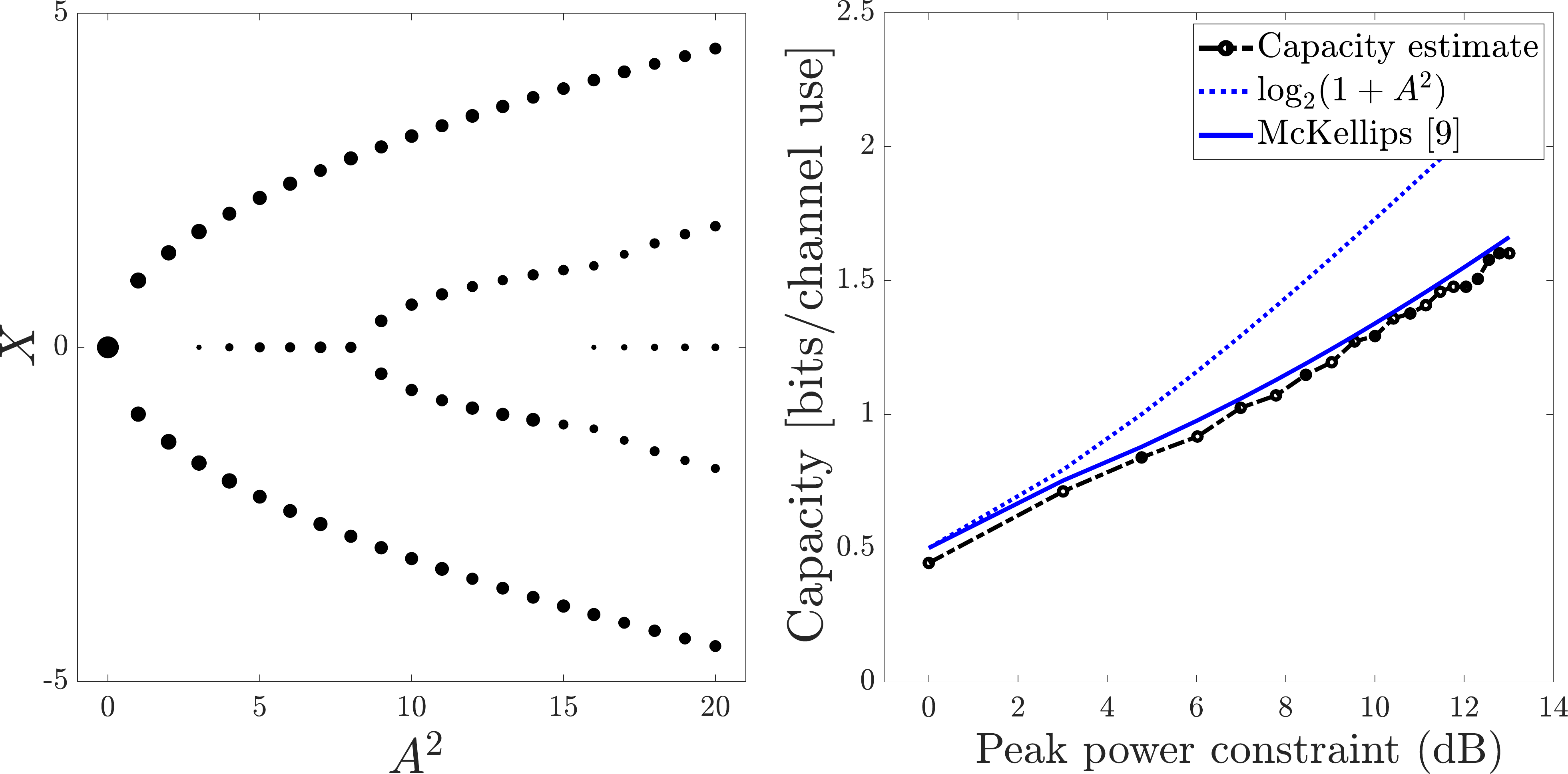}
	\caption{AWGN scalar peak-power constrained channel: a) capacity-achieving distribution learned by CORTICAL, the marker's radius is proportional to the PMF; b) capacity estimate and comparisons.}
	\label{fig:bifurcation}
\end{figure} 
For convenience of comparison and as a proof of concept, we focus on the scalar $d=1$ and $d=2$ channels. 
The first findings on the capacity-achieving discrete distributions were reported in \cite{Smith1971}. For a scalar channel, it was shown in \cite{Sharma2010} that the input has alphabet $\{-A,A\}$ with equiprobable values if $0<A \lessapprox 1.6$, while it has ternary alphabet $\{-A,0,A\}$ if $1.6 \lessapprox A \lessapprox 2.8$.
Those results are confirmed by CORTICAL which is capable of both understanding the discrete nature of the input and learning the support and probability mass function (PMF) for any value of $A$. Notice that no hypothesis on the input distribution is provided during training. Fig. \ref{fig:bifurcation}a reports the capacity-achieving input distribution as a function of the peak-power constraint $A^2$. The figure illustrates the typical bifurcation structure of the distribution. Fig. \ref{fig:bifurcation}b shows the channel capacity estimated by CORTICAL with \eqref{eq:capacity} and compares it with the upper bounds known in the literature.  

\begin{figure}
	\centering
	\includegraphics[scale=0.28]{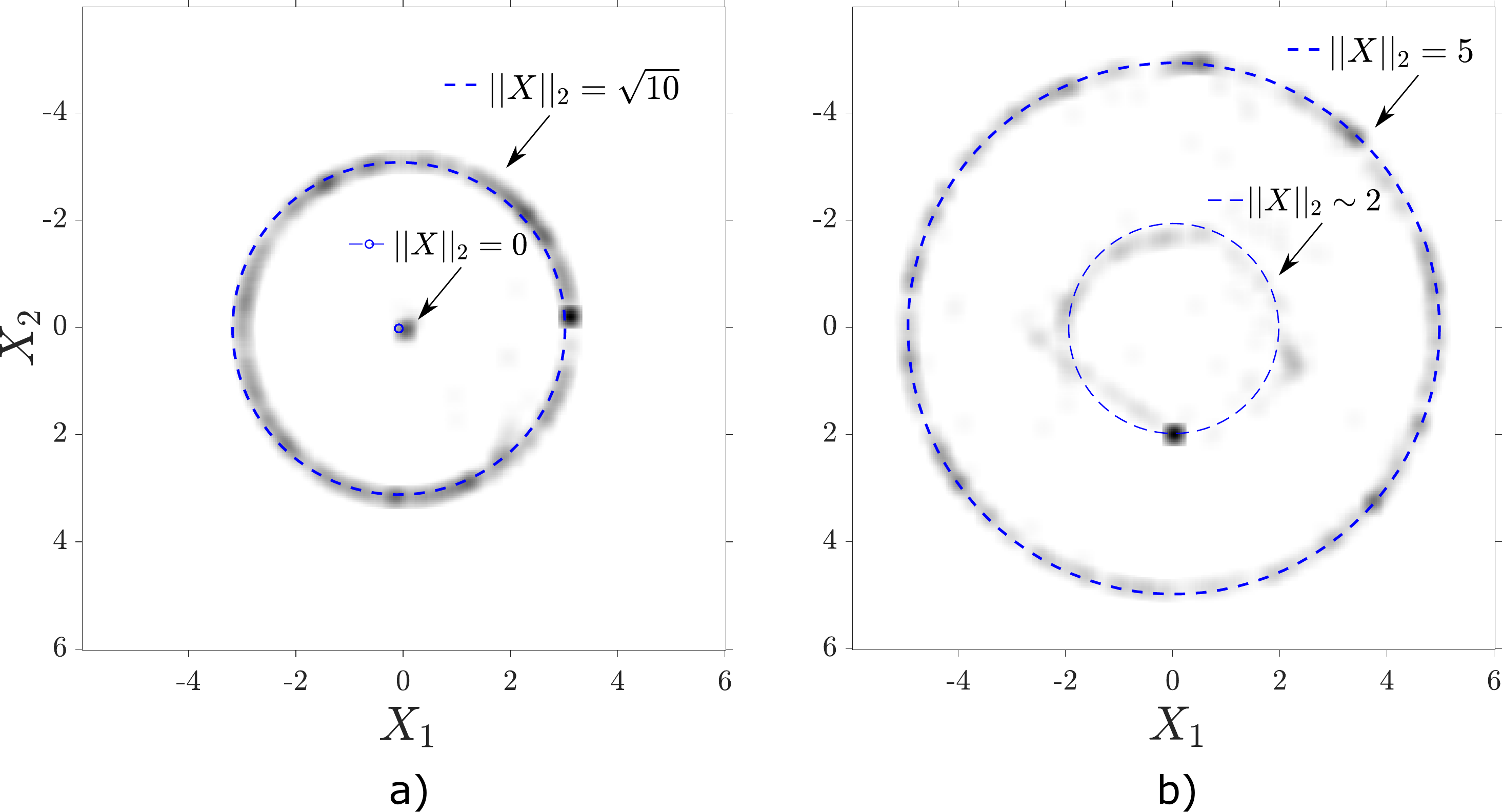}
	\caption{AWGN $d=2$ channel input distributions learned by CORTICAL under a peak-power constraint: a) $A=\sqrt{10}$; b) $A=5$.}
	\label{fig:heatmaps}
\end{figure} 

\begin{figure}
	\centering
	\includegraphics[scale=0.28]{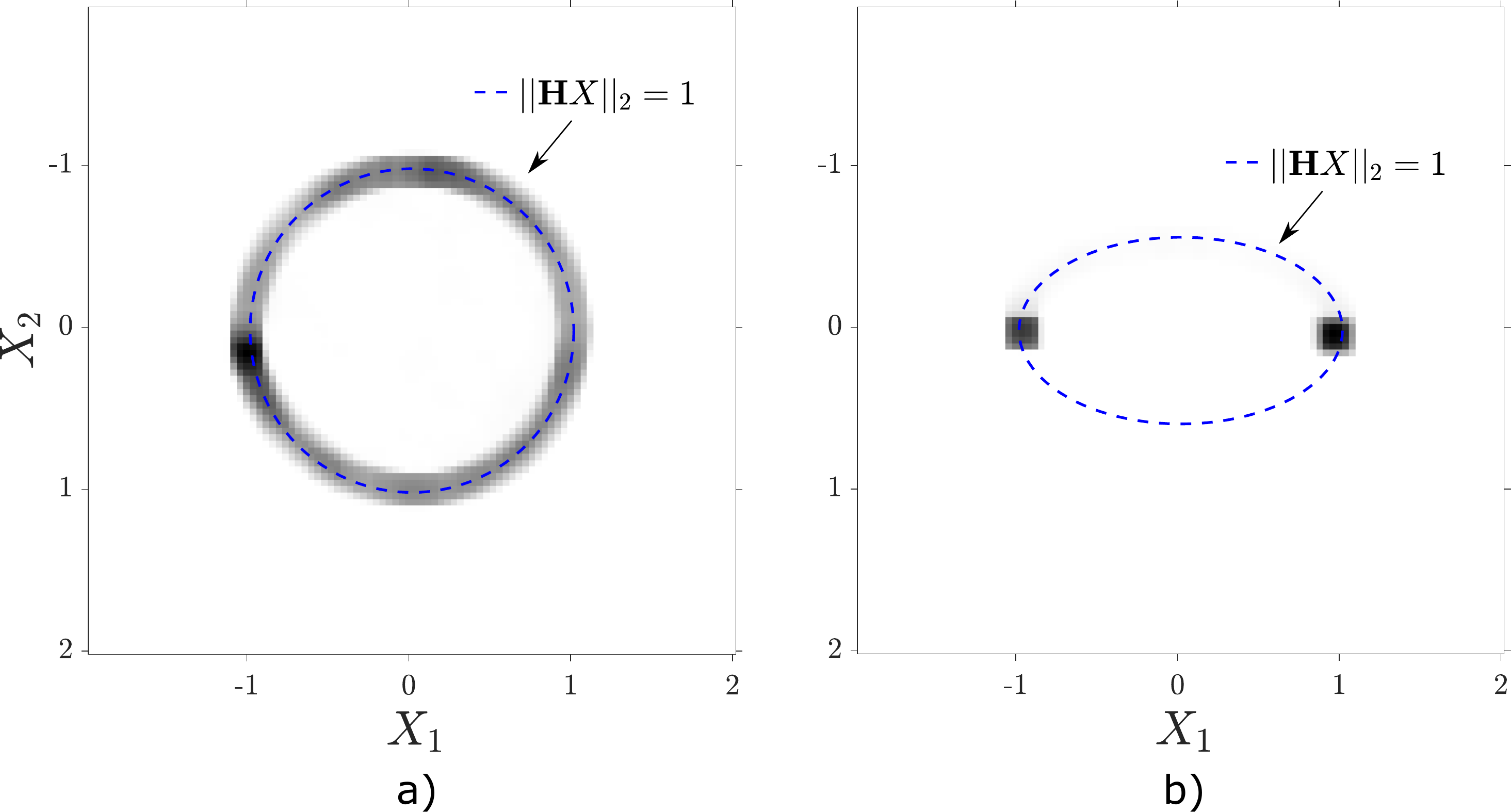}
	\caption{MIMO channel input distributions learned by CORTICAL for different channels $\mathbf{H}$: a) $r_2=1$; b) $r_2=3$. Locus of points satisfying $||\mathbf{H}X||_2= 1$ is also shown.}
	\label{fig:MIMO}
\end{figure} 
Similar considerations can be made for the $d=2$ channel under peak-power constraints \cite{Rassouli2016}, and in general for the vector Gaussian channel of dimension $d$ \cite{Dytso2019}. 
The case $d=2$ is considered in Fig. \ref{fig:heatmaps}a ($A=\sqrt{10}$) and Fig. \ref{fig:heatmaps}b ($A=5$) where CORTICAL learns optimal bi-dimensional distributions, matching the results in \cite{Rassouli2016}. In this case the amplitude is discrete. 

We now consider the MIMO channel where the analytical characterization of the capacity-achieving distribution under a peak-power constraint remains mostly an open problem \cite{Dytso2019-MIMO}. We analyze the particular case of $d=2$: 
\begin{equation}
C(\mathbf{H},r) = \sup_{p_{X}(\mathbf{x}): ||\mathbf{H}X||_2\leq r} I(X;\mathbf{H}X+N),
\end{equation}
where $N \sim \mathcal{N}(0,\mathbf{I}_2)$ and $\mathbf{H} \in \mathbb{R}^{2\times 2}$ is a given MIMO channel matrix known to both transmitter and receiver. We impose $r=1$, and we also impose a diagonal structure on $\mathbf{H}=\text{diag}(1,r_2)$ without loss of generality since diagonalization of the system can be performed. We study two cases: $r_2 = 1$, which is equivalent to a $d=2$ Gaussian channel with unitary peak-power constraint; and $r_2=3$, which forces an elliptical peak-power constraint. The former set-up produces a discrete input distribution in the magnitude and a continuous uniform phase, as shown in Fig. \ref{fig:MIMO}a. The latter case produces a binary distribution, as shown in Fig. \ref{fig:MIMO}b. To the best of our knowledge, no analytical results are available for $1<r_2<2$ and thus CORTICAL offers a guiding tool for the identification of capacity-achieving distributions. 

\subsection{Additive Non-Gaussian Channels}
The nature of the capacity-achieving input distribution remains an open challenge for channels affected by additive non-Gaussian noise. In fact, it is known that under an average power constraint, the capacity-achieving input distribution is discrete and the AWGN channel is the only exception \cite{Fahs2012}. Results concerning the number of mass points of the discrete optimal input have been obtained in \cite{Tchamkerten2004,Das2000}. If the transmitter is subject to an average power constraint, the support is bounded or unbounded depending on the decay rate (slower or faster, respectively) of the noise PDF tail.

\begin{comment}
For the purposes of this section, we consider a scalar additive linear channel with exponential noise under an average power input constraint. In particular, we consider a symmetric exponential noise PDF $p_N(n) = e^{-2|n|}$ and show that the capacity-achieving distribution is discrete and has bounded support.
\end{comment}
\begin{figure}
	\centering
	\includegraphics[scale=0.34]{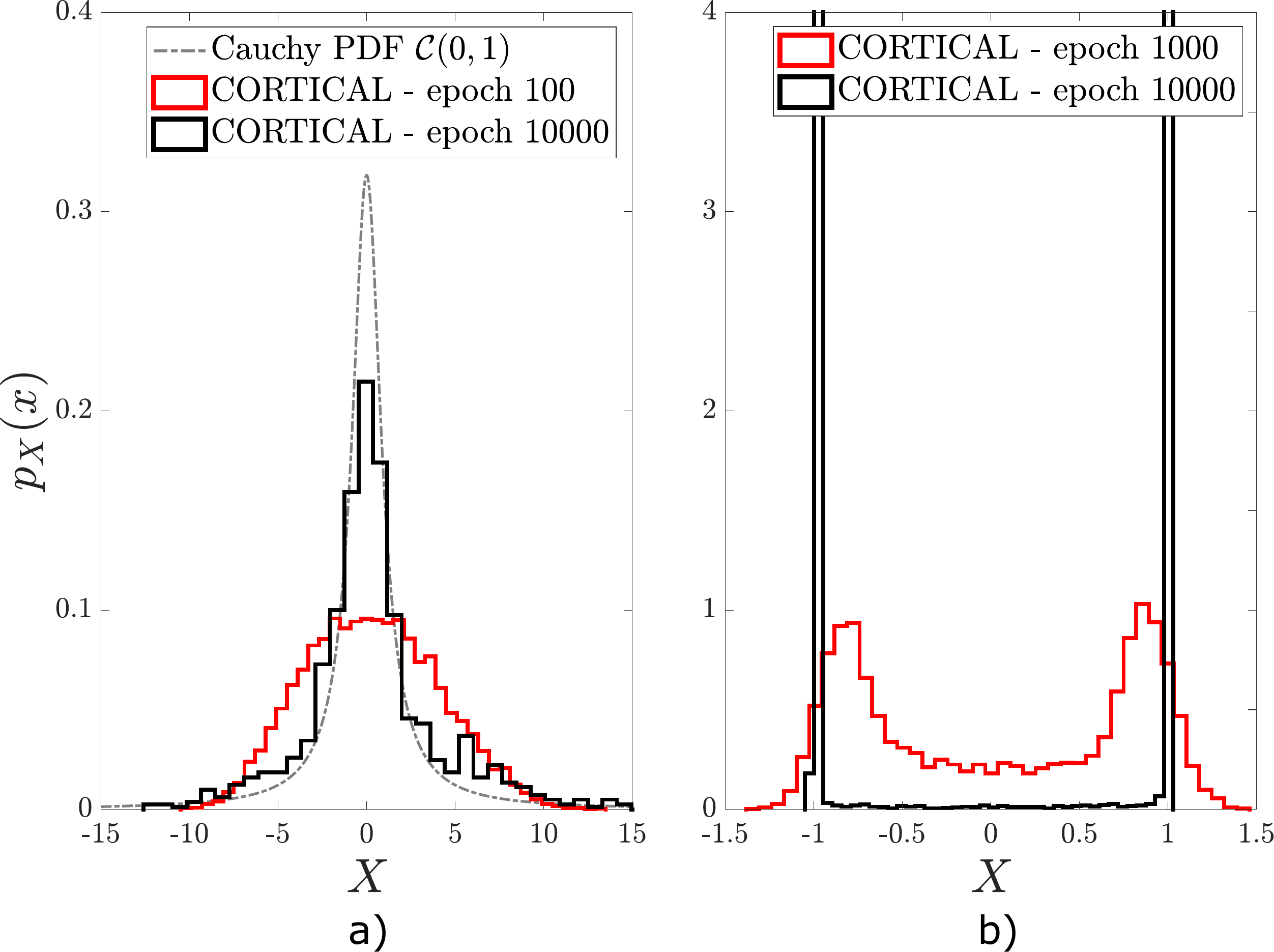}
	\caption{AICN channel input distributions learned by CORTICAL at different training steps under: a) logarithmic power constraint; b) peak-power constraint.}
	\label{fig:cauchy_subplot}
\end{figure} 
In this section, we consider a scalar additive independent Cauchy noise (AICN) channel with scale factor $\gamma$, under a specific type of logarithmic power constraint \cite{Fahs2014}. In particular, given the Cauchy noise PDF $\mathcal{C}(0,\gamma)$
\begin{equation}
    p_N(n) = \frac{1}{\pi \gamma}\frac{1}{1+\bigl(\frac{n}{\gamma}\bigr)^2},
\end{equation}
we are interested in showing that CORTICAL learns the capacity-achieving distribution that solves
\begin{equation}
C(A,\gamma) = \sup_{p_X(\mathbf{x}): \mathbb{E}\bigl[\log\bigl(\bigl(\frac{A+\gamma}{A}\bigr)^2+\bigl(\frac{X}{A}\bigr)^2\bigr)\bigr]\leq \log(4)} I(X;Y),
\end{equation}
for a given $A\geq \gamma$. From \cite{Fahs2014}, it is known that under such power constraint the channel capacity is $C(A,\gamma)=\log(A/\gamma)$ and the optimal input distribution is continuous with Cauchy PDF $\mathcal{C}(0,A-\gamma)$. For illustration purposes, we study the case of $\gamma=1$ and $A=2$ and report in Fig. \ref{fig:cauchy_subplot}a the input distribution obtained by CORTICAL after $100$ and $10000$ training steps. For the same channel, we also investigate the capacity-achieving distribution under a unitary peak-power constraint. Fig. \ref{fig:cauchy_subplot}b shows that the capacity achieving distribution is binary.

\subsection{Fading Channels}
As the last representative scenario, we study the Rayleigh fading channel subject to an average power constraint $P$ with input-output relation given by
\begin{equation}
    Y = \alpha X + N,
\end{equation}
where $\alpha$ and $N$ are independent circular complex Gaussian random variables, so that the amplitude of $\alpha$ is Rayleigh-distributed.
It is easy to prove that an equivalent model for deriving the distribution of the amplitude of the signal achieving-capacity is obtained by defining $U=|X|\sigma_{\alpha}/\sigma_N$ and $V=|Y|^2/\sigma_N^2$, which are non-negative random variables such that $\mathbb{E}[U^2]\leq P\frac{\sigma_{\alpha}^2}{\sigma_N^2}\triangleq a$ and whose conditional distribution is
\begin{equation}
    p_{V|U}(v|u) = \frac{1}{1+u^2} \exp \biggl(-\frac{v}{1+u^2}\biggr).
\end{equation}
A simplified expression for the conditional pdf is 
\begin{equation}
\label{eq:rayleigh_channel}
    p_{V|S}(v|s) = s \cdot \exp (-sv)
\end{equation}
where $S = 1/(1+U^2) \in (0,1]$ and $\mathbb{E}[1/S-1]\leq a$.
From \cite{Faycal2001}, it is known that the optimal input signal amplitude $S$ (and thus $U$) is discrete and possesses an accumulation point at $S=1$ ($U=0)$. We use CORTICAL to learn the capacity-achieving distribution and verify that it matches \eqref{eq:rayleigh_channel} for the case $a=1$ reported in \cite{Faycal2001}. 
\begin{figure}
	\centering
	\includegraphics[scale=0.20]{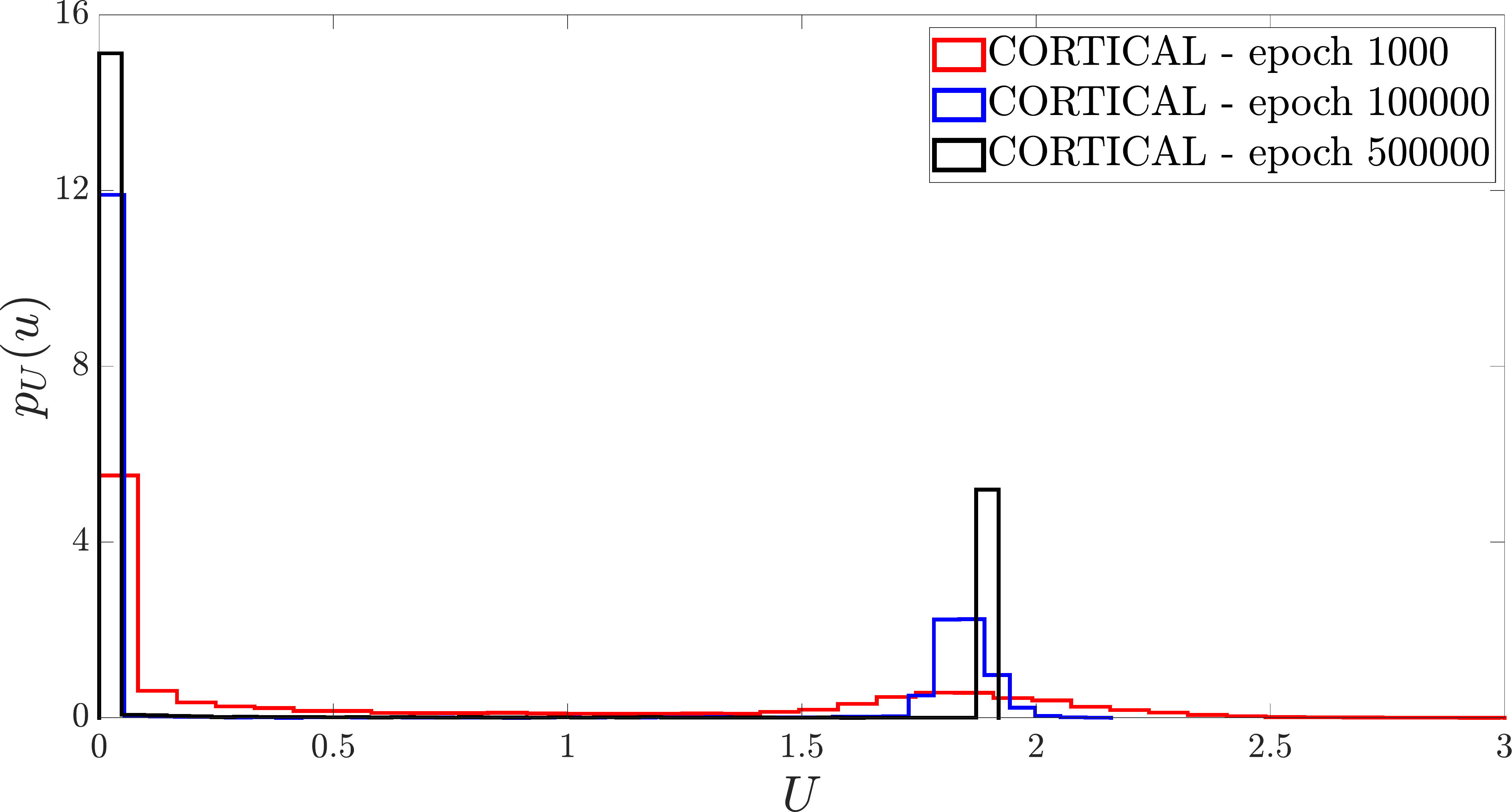}
	\caption{Optimal input $U$ learned by CORTICAL at different training steps.}
	\label{fig:rayleigh}
\end{figure} 
Fig. \ref{fig:rayleigh} illustrates the evolution of the input distribution for different CORTICAL training steps. Also in this scenario, the generator learns the discrete nature of the input distribution and the values of the PMF. %as predicted in \cite{Faycal2001}. 

\begin{comment}
\subsubsection{Poisson noise channel}
Discrete-time memoryless Poisson noise channel
\begin{equation}
    P_{Y|X}(y|x) = \frac{1}{y!}(ax+\lambda)^y e^{-(ax+\lambda)}, x\geq 0, y\in \mathbb{N}
\end{equation}
where $a>0$ is the scaling factor and $\lambda \geq 0$ is the dark-current term. $Y=H(X)$ randomly transforms the channel input $X$ into a discrete channel output $Y$.
There are two sources of noise in this channel: 1. the probabilistic nature in which spikes are generated at the transmitter, and 2. the extraneous spikes due to background emission (dark current). We focus only on the first term.

\subsubsection{Finite input-alphabet}
Discrete input case, codes..
\end{comment}
\section{Conclusions}
\label{sec:conclusions}
In this paper, we have presented CORTICAL, a novel deep learning-based framework that learns the capacity-achieving distribution for any discrete-time continuous memoryless channel. The proposed approach consists of a cooperative max-max game between two neural networks. Non-Shannon channel settings have validated the algorithm. The proposed approach offers a novel tool for studying the channel capacity in non-trivial channels and paves the way for the holistic and data-driven analysis/design of communication systems. 
\section{Acknowledgement}
The authors would like to thank Nir Weinberger and Alex Dytso for the valuable discussions about capacity-achieving distributions provided while Nunzio A. Letizia was visiting Princeton University.
\bibliographystyle{IEEEtran}
\bibliography{IEEEabrv,biblio}

\begin{thebibliography}{10}
\providecommand{\url}[1]{#1}
\csname url@samestyle\endcsname
\providecommand{\newblock}{\relax}
\providecommand{\bibinfo}[2]{#2}
\providecommand{\BIBentrySTDinterwordspacing}{\spaceskip=0pt\relax}
\providecommand{\BIBentryALTinterwordstretchfactor}{4}
\providecommand{\BIBentryALTinterwordspacing}{\spaceskip=\fontdimen2\font plus
\BIBentryALTinterwordstretchfactor\fontdimen3\font minus
  \fontdimen4\font\relax}
\providecommand{\BIBforeignlanguage}[2]{{%
\expandafter\ifx\csname l@#1\endcsname\relax
\typeout{** WARNING: IEEEtran.bst: No hyphenation pattern has been}%
\typeout{** loaded for the language `#1'. Using the pattern for}%
\typeout{** the default language instead.}%
\else
\language=\csname l@#1\endcsname
\fi
#2}}
\providecommand{\BIBdecl}{\relax}
\BIBdecl

\bibitem{Oshea2017}
T.~{O'Shea} and J.~{Hoydis}, ``An introduction to deep learning for the
  physical layer,'' \emph{IEEE Trans. on Cognitive Communications and
  Networking}, vol.~3, no.~4, pp. 563--575, Dec 2017.

\bibitem{Ye2017}
H.~Ye, G.~Y. Li, and B.-H. Juang, ``Power of deep learning for channel
  estimation and signal detection in {OFDM} systems,'' \emph{IEEE Wireless
  Communications Letters}, vol.~7, no.~1, pp. 114--117, 2018.

\bibitem{MIND2022}
A.~M. Tonello and N.~A. Letizia, ``Mind: Maximum mutual information based
  neural decoder,'' \emph{IEEE Communications Letters}, vol.~26, no.~12, pp.
  2954--2958, 2022.

\bibitem{Ye2018}
H.~Ye, G.~Y. Li, B.-H.~F. Juang, and K.~Sivanesan, ``Channel agnostic
  end-to-end learning based communication systems with conditional gan,'' in
  \emph{2018 IEEE Globecom Workshops (GC Wkshps)}, 2018, pp. 1--5.

\bibitem{Soltani2019}
M.~Soltani, V.~Pourahmadi, A.~Mirzaei, and H.~Sheikhzadeh, ``Deep
  learning-based channel estimation,'' \emph{IEEE Communications Letters},
  vol.~23, no.~4, pp. 652--655, 2019.

\bibitem{Aharoni2020}
Z.~Aharoni, D.~Tsur, Z.~Goldfeld, and H.~H. Permuter, ``Capacity of continuous
  channels with memory via directed information neural estimator,'' in
  \emph{Proc. of the 2020 IEEE International Symposium on Inf. Theory (ISIT)},
  2020, pp. 2014--2019.

\bibitem{Letizia2021}
N.~A. Letizia and A.~M. Tonello, ``Capacity-driven autoencoders for
  communications,'' \emph{IEEE Open Journal of the Communications Society},
  vol.~2, pp. 1366--1378, 2021.

\bibitem{Farhad2022}
F.~Mirkarimi, S.~Rini, and N.~Farsad, ``Neural capacity estimators: How
  reliable are they?'' in \emph{Proc. of the ICC 2022 - IEEE International
  Conference on Communications}, 2022, pp. 3868--3873.

\bibitem{CuttingPlane}
J.~Huang and S.~Meyn, ``Characterization and computation of optimal
  distributions for channel coding,'' \emph{IEEE Trans. on Inf. Theory},
  vol.~51, no.~7, pp. 2336--2351, 2005.

\bibitem{McKellips2004}
A.~McKellips, ``Simple tight bounds on capacity for the peak-limited
  discrete-time channel,'' in \emph{Proc. of the 2004 IEEE International
  Symposium on Inf. Theory}, 2004, pp. 348--348.

\bibitem{Smith1971}
J.~G. Smith, ``The information capacity of amplitude- and variance-constrained
  scalar {G}aussian channels,'' \emph{Inf. Control.}, vol.~18, pp. 203--219,
  1971.

\bibitem{Dytso2020}
A.~Dytso, S.~Yagli, H.~V. Poor, and S.~Shamai~Shitz, ``The capacity achieving
  distribution for the amplitude constrained additive {G}aussian channel: An
  upper bound on the number of mass points,'' \emph{IEEE Trans. on Inf.
  Theory}, vol.~66, no.~4, pp. 2006--2022, 2020.

\bibitem{BlahutArimoto}
R.~Blahut, ``Computation of channel capacity and rate-distortion functions,''
  \emph{IEEE Trans. on Inf. Theory}, vol.~18, no.~4, pp. 460--473, 1972.

\bibitem{Dauwels2005}
J.~Dauwels, ``Numerical computation of the capacity of continuous memoryless
  channels,'' \emph{Proc. of the 26th Symposium on Inf. Theory in the Benelux},
  2005.

\bibitem{Goodfellow2014}
I.~Goodfellow, J.~Pouget-Abadie, M.~Mirza, B.~Xu, D.~Warde-Farley, S.~Ozair,
  A.~Courville, and Y.~Bengio, ``Generative adversarial nets,'' in
  \emph{Advances in Neural Information Processing Systems}, vol.~27, 2014.

\bibitem{CORTICAL_github}
N.~A. Letizia and A.~M. Tonello, ``\uppercase{CORTICAL} cooperative channel
  capacity learning,'' \url{https://github.com/tonellolab/CORTICAL}, 2022.

\bibitem{Faycal2001}
I.~Abou-Faycal, M.~Trott, and S.~Shamai, ``The capacity of discrete-time
  memoryless {R}ayleigh-fading channels,'' \emph{IEEE Trans. on Inf. Theory},
  vol.~47, no.~4, pp. 1290--1301, 2001.

\bibitem{Tchamkerten2004}
A.~Tchamkerten, ``On the discreteness of capacity-achieving distributions,''
  \emph{IEEE Trans. on Inf. Theory}, vol.~50, no.~11, pp. 2773--8, 2004.

\bibitem{Fahs2014}
J.~Fahs and I.~Abou-Faycal, ``A {C}auchy input achieves the capacity of a
  {C}auchy channel under a logarithmic constraint,'' in \emph{Proc. of the 2014
  IEEE International Symposium on Inf. Theory}, 2014, pp. 3077--3081.

\bibitem{Rassouli2016}
B.~Rassouli and B.~Clerckx, ``On the capacity of vector {G}aussian channels
  with bounded inputs,'' \emph{IEEE Trans. on Inf. Theory}, vol.~62, no.~12,
  pp. 6884--6903, 2016.

\bibitem{Sharma2010}
N.~Sharma and S.~Shamai, ``Transition points in the capacity-achieving
  distribution for free-space optical intensity channels,'' \emph{Proceedings
  of the 2010 IEEE Inf. Theory Workshop}, pp. 1--5, 2010.

\bibitem{Dytso2019}
A.~Dytso, M.~Al, H.~V. Poor, and S.~Shamai~Shitz, ``On the capacity of the peak
  power constrained vector {G}aussian channel: An estimation theoretic
  perspective,'' \emph{IEEE Trans. on Inf. Theory}, vol.~65, no.~6, pp.
  3907--3921, 2019.

\bibitem{Dytso2019-MIMO}
A.~Dytso, M.~Goldenbaum, H.~V. Poor, and S.~Shamai~(Shitz), ``Amplitude
  constrained \uppercase{MIMO} channels: Properties of optimal input
  distributions and bounds on the capacity,'' \emph{Entropy}, vol.~21, no.~2,
  2019.

\bibitem{Fahs2012}
J.~Fahs, N.~Ajeeb, and I.~Abou-Faycal, ``The capacity of average power
  constrained additive non-{G}aussian noise channels,'' in \emph{Proc. of the
  19th International Conference on Telecommunications}, 2012, pp. 1--6.

\bibitem{Das2000}
A.~Das, ``Capacity-achieving distributions for non-{G}aussian additive noise
  channels,'' \emph{Proc. of the 2000 IEEE International Symposium on Inf.
  Theory}, p. 432, 2000.

\end{thebibliography}

\end{document}